\documentclass[10pt]{article}
\usepackage[explicit]{titlesec}
\usepackage{hyphenat}
\usepackage{ragged2e}
\usepackage{enumitem}
\usepackage{amssymb}
\usepackage{tcolorbox}
\usepackage{scalerel,amssymb}
\usepackage{mathtools}
\usepackage[makeroom]{cancel}
\usepackage{multirow}
\usepackage{booktabs}
\usepackage{multicol}
\usepackage[utf8]{inputenc}
\usepackage{graphicx}
\usepackage{soul}
\setul{.6pt}{.4pt}
\usepackage{xcolor}
\usepackage{stackrel} 

\usepackage{framed}
\usepackage{lipsum}

\usepackage{amsmath}

\usepackage[labelfont={footnotesize,bf} , textfont=footnotesize]{caption}
\makeatletter
\renewcommand\tagform@[1]{\maketag@@@ {\ignorespaces {\footnotesize{\textbf{Equation}}} #1.\unskip \@@italiccorr }}
\makeatother
\usepackage{framed}

\usepackage{geometry}
\usepackage{fancyhdr}
\makeatletter
\newcommand\sixteen{\@setfontsize\sixteen{17pt}{6}}
\renewcommand{\maketitle}{\bgroup\setlength{\parindent}
\begin{flushleft}
\sixteen\bfseries \@title
\medskip
\end{flushleft}
\textit{\@author}
\egroup}
\makeatother
\usepackage{float}

\begin{document}

\title{ECON4200 Senior Seminar in Economics and Finance}
\author{John T. H. Wong, Matthias Hei Man, and Alex Li Cheuk Hung}
\date{ }

\begin{center}

\thispagestyle{empty}

   \Large  \text{ECON 4200 Senior Seminar in Economics and Finance} \\ 
   \Large \text{Population and Technological Growth: Evidence from \emph{Roe v. Wade}} \\
\end{center}

\begin{center} 
        John T. H. Wong, 3035592600  \\ 
        Matthias Hei Man, 3035552375  \\
        Alex Li Cheuk Hung, 3035492288  \\

\end{center}

\section{Introduction}

Some economists have argued that a greater population base causes higher technological growth and therefore higher per capita income, whether that is due to network effects such as intellectual contact and specialization spurring innovation or due to the need to sustain a larger population. Meanwhile, others have argued that per capita income decreases as population grows, due to competition over a fixed set of resources and greater dependency. Furthermore, it is commonly known that cross-sectional empirical evidence shows countries with higher population growth having lower income. This paper will attempt to validate the first position by providing evidence that a greater population leads to more technological growth in the form of patent production.

We find that for cohorts born between 1931-1984, a higher starting population at birth is correlated with higher patents per thousand residents between 1996-2012. In order to rule out the endogeneity of fertility decisions and estimate the causal effect of cohort births\footnote{We use cohort starting population and cohort births interchangeably throughout this paper.} on patent production, we exploit the heterogeneous impact of the US Supreme Court's ruling on \emph{Roe v. Wade}, which ruled most abortion restrictions unconstitutional. Our identifying assumption is that states which had not liberalized their abortion laws prior to \emph{Roe} would experience a negative birth shock of greater proportion than states which had undergone pre-\emph{Roe} reforms. We estimate the difference-in-difference in births and use estimated births as an exogenous treatment variable to predict patents per capita. Our results show that one standard deviation increase in cohort's starting population (70,608 births) increases per capita patents by 0.23 (which is 24 percent of the outcome's standard deviation). These results suggest that at the margins, increasing fertility can increase patent production. Insofar as patent production is a sufficient proxy for technological growth, increasing births has a positive impact on technological growth.

Our paper builds on two fields of research: the first is theories on the relationship between population and innovation, whose contributors include Michael Kremer, Simon Kuznets, and David Weil. Second, we add to the study of determinants of innovation in the US by Bell et al. Here, we would also like to acknowledge Bell et al. for making the data set of their study open-source and allowing this instance of alternative use. We should also note that although our research uses changes in abortion policy as an exogenous change in births, this paper and its results \emph{do not} pertain to the issue of abortion itself. Insofar as we prove that fertility has a positive impact on technological growth (which we will argue is the case), we establish that states have good reason to promote births. We are silent on the optimality of abortion policy as a natalist tool for increasing technological growth.

Before we proceed, we should also take stock of the limitations of our paper. First, as it will become clear in Section 4, our use of \emph{Roe} as an exogenous change in births is subject to limitations. Although states with pre-\emph{Roe} bans on abortion see births decrease, this drop is not statistically significant, making our IV quite weak. Our difference-in-difference estimations of patents per capita (without using cohort births as a treatment variable) also contradict our hypothesis, which we think is most likely due to confounding factors that may have occurred in non-reform states after \emph{Roe}, which we could not control for due to our limited research capacity.

Second, our results (insofar as they are valid) are estimated with instrumental variables, which make them estimated local average treatment effects (LATE). We can only estimate the effect of births on innovation for individuals of cohorts who became able to undergo abortions as a result of the repeal of abortion laws following the \emph{Roe} ruling. This means, we cannot know the effect of births on patent production for individuals who would have sought abortions regardless of the procedure's legality in their home state (by travelling out-of-state), or individuals who would not have sought abortions regardless, i.e. all eventual parents after \emph{Roe}. Furthermore, we can only estimate LATE through the compliers' influence on their cohorts' average level of births and patents.

Finally, it is also beyond our research capacity to account for the heterogeneity in patent utility across patents. This could be done by merging patents per capita of a cohort to how much the cohort's patents were cited. However, our data set on patent outcomes do not track patent utility by cohort and state. To our knowledge, the US Patent and Trademark Office (USPTO) also does not release public data of such kind. As a result, we can only assume all patents granted are equal in their contribution to technological growth, however flawed that assumption might be.

Our paper will proceed as follows. Section 2 discusses the background of our research. We will review the relevant theoretical literature on population, technological growth, and income. We also will discuss what determines an individual's chance of becoming patent holders in the US context. Section 3 describes the various data sets we combine and use, and estimate the correlation between births and patents. Section 4 covers the institutional setting of abortion laws in the US and attempts to justify the use of the \emph{Roe} ruling as an exogenous shock in births. In Section 5, we will describe our methodology for and results from estimating the causal effect of births on patents. We will also analyze where our methodology falls short and discuss its major limitations. Section 6 concludes.

\vspace{.5cm}

\section{Background}

\subsection{Theoretical Framework and Literature}

Our hypothesis that population and technology have a positive relationship follows Kremer's model and evidence on this subject. Kremer offers two main views. First, he co-opts the “Malthusian assumption that technology limits population” (Kremer 1993, 681) to argue that high population forces the adoption of “new” technology to replace “old” technology (i.e. technology insufficient for supporting a given level of population). Research productivity, under this view, would depend on the level of existing population and we should see proportionality between the growth of these two variables (Kremer 1993, 681-682). To support this, Kremer shows that eras with greater population bases also have higher population growth rates. In other words, because of the positive effect of population base on technological growth, humanity has been able to afford super-exponential population growth. Second, Kremer rejects the view that subsequent rises in income would have reduced efforts to invent new technologies (Kremer 1993, 684). Instead, he argues that research productivity depends positively on income (Kremer 1993, 687). That high population without income is insufficient for achieving technological growth also explains why densely populated countries such as China had (as of 1990) low research productivity. 

Kremer's model and results contradict the general view that population growth reduces per capita income. Thomas Malthus has argued that larger populations will eventually fail due to the inadequacy of resources. Kremer's arguments are also contrary to economic growth models such as the Solow and Harrod-Domar models, which both predict that societies with higher population growth will see lower levels of per capita income (Williamson 2014, 222-5; 248-9; Ray 1998, 51-6). More recently, Weil has argued that as the populations of developing countries age, increasing fertility could actually lead to less per capita income in the short-term as dependency increases (Weil 1999, 253).\footnote{There is also a strand of literature on demographic transition which argues that as economic development improves, income increases, and child mortality decreases, households require less children as an investment into old-age security and therefore desire less children. However, here, the focus is on the relationship from income to population, rather than from population to income. We focus on the latter as our main interest is in determining what drives technological (and therefore economic) growth. For more, see Robert J. Barro and Gary S. Becker. "Fertility Choice in a Model of Economic Growth." \emph{Econometrica} 57 (1989): pp. 481-501.} Galor and Weil (1999) have argued that even if Kremer is correct that a greater population base leads to more technological growth, this growth will subsequently reward investments in human capital that will (i) have a greater role driving subsequent technological growth and therefore (ii) lead households to prioritize the quality of children over quantity, explaining low levels of population growth in developed economies. Our task is to argue that the population-induced growth is still significant even in the context of a developed economy. However, it is beyond the scope of this paper to compare the size of population effects to the size of human capital investment's effect on innovation.

Separately, Kuznets has argued that productivity per capita, and by extension innovation per capita, should increase with a larger population. For example, consider country A with population size 10 and country B with a larger population, say 20. All else equal, the productivity per capita of B should be greater than A because higher population density allows for greater division of labor and specialization, and the “possibility of more intensive intellectual contact" (Kuznets 1960, 325-327). Furthermore, Kuznets notes that growing population increases the size of the market and improves its responsiveness to new goods (a proxy of innovation) (Kuznets 1960, 334-347). In other words, a greater population acts as a greater financial incentive for productivity growth. Such network effects suggest that population growth and patent production is non-linear (i.e. greater populations see higher patents per capita and exponentially more patents), which is precisely the hypothesis we aim to prove in the US context.

Bell et al. provides a recent examination of innovation in modern US. The authors evaluate individuals who end up as innovators in the US and delve into the backgrounds of these innovators (defined as those who produce patents) in terms of race, gender, income, and other birth characteristics to obtain explanatory factors. For example, they highlight how higher percentile income is linked to an exponentially higher number of inventors per thousand, showing that “rates of innovation rise by 22 percent between the 95th percentile (\$193,322) and 99th percentile (\$420,028) of the parental income distribution” (Bell et al. 2018, 12). We illustrate this relationship by plotting the number of inventors per capita across quintiles in Figure 1. Race or ethnicity is another factor highlighted, with evidence that “1.6 per 1,000 white children and 3.3 per 1,000 Asian children who attend NYC public schools between grades 3-8 become inventors,” relative to only 0.5 for Black Children and 0.2 for Hispanics (Bell et al. 2018, 13-14). Finally, there is also a gap in innovation between genders, with men more likely to become inventors. In 1980, the fraction of female inventors is only around 20 percent, and Bell et al. estimate that, given the current rate of convergence estimated from linear regression, it would approximately take until 2100 to reach a 50 percent female share (Bell et al. 2018, 14). \vspace{.3cm}

\begin{figure}[h]
\centering
\caption{}
\vspace{.5cm}
\includegraphics[width=15cm]{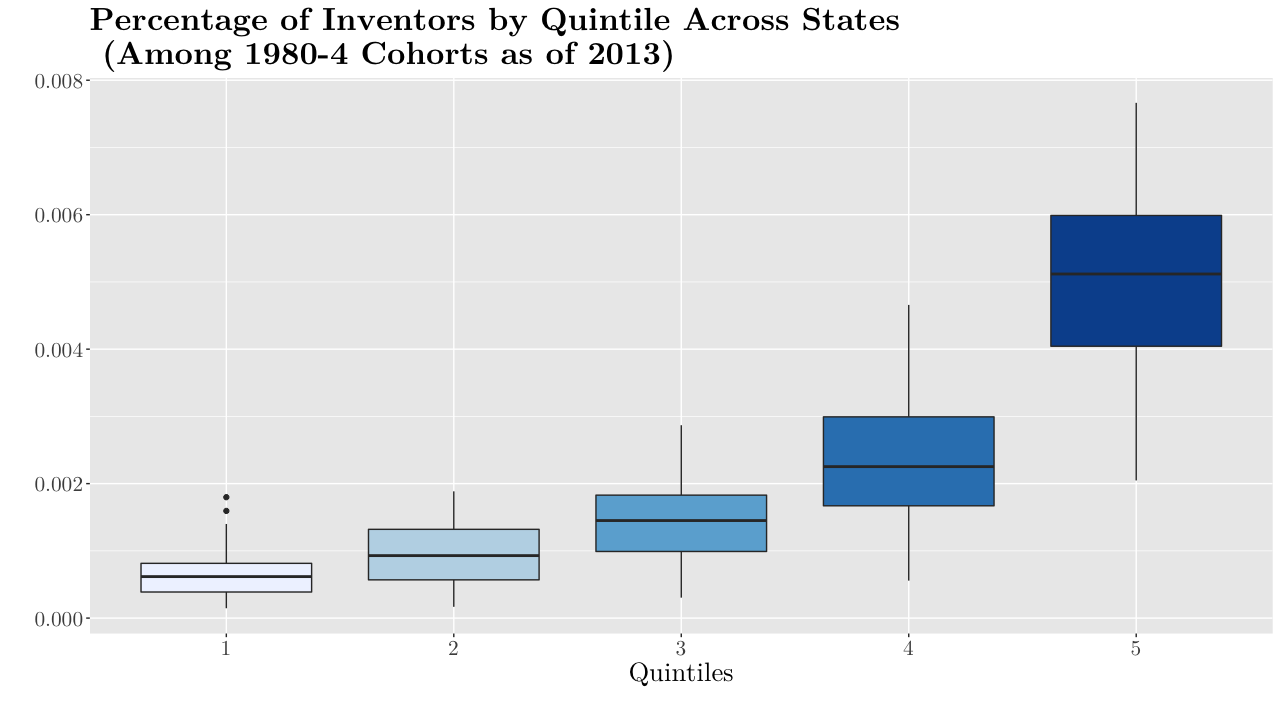}
\end{figure}
\vspace{.5cm}

Bell et al. also highlight the importance of spatial determinants. Evidence on spatial determinants of innovation shows that commuting distances have significant impacts on patent productivity–both in terms of commuting during childhood and adulthood. This makes sense as longer commuting times decrease time available to be productive. Research by Xiao et al. shows that the distance between an individual’s home and their workplace versus their patent productivity is negatively correlated. An increase of ten kilometers in the worker’s commuting distance corresponds to a “5\% decrease in patents per inventor–firm pair per year and an even greater 7 percent decrease in patent quality” (Xiao et al. 2021, 1; 13). Bell et al. further show that neighbourhoods where children grow up in are also significant. Bell et al. argue that children growing up in areas where innovation activity is higher are more likely to become innovators themselves thanks to exposure to innovation. Their regression shows that a one standard deviation increase in a neighbourhood’s (or community zone’s) patent rate corresponds with an increase in the fraction of children who become inventors, having lived in the same neighbourhood, by 28.5 percent (Bell et al. 2018, 24-28). The results of these papers highlight that financial incentives are not the only motivating factor in patent production, as exposure and geography itself from birth, childhood, to adulthood can impact an individual’s probability of becoming an inventor. 

\vspace{.5cm}

\section{Data and Baseline Results}

\subsection{Data}

In order to estimate the effect of births on innovation, we assemble, construct, and merge several data sets. We use Bell et al.’s open-source panel data on patent outcomes. The panel data tracks patent outcomes across three units of observations. First, patent outcomes are tracked for each cohort born between 1916-1984; as patent outcomes are clustered by cohort (rather than at an individual level), cohorts are also the main unit of observation. Second, patent outcomes for each cohort are reported once per year, between 1996-2014. For example, we separately know how many patents the 1970 cohort were granted in 1996, 1997, and so forth. The data set only includes cohorts aged 25-80 each year. We track outcomes by year because a given cohort’s propensity to produce patents is highly dependent on the cohort’s age (see Figure 2). Finally, patent outcomes for each cohort in each year are reported separately for each US state. To extend on the previous example, one row in the data set tells us how many patents were granted in 1996 to the 1970 cohort in California (versus, say, in Oregon). Patents granted in a year refer to patents that were applied in that year and subsequently granted even if the latter occurs in later years. Because the data only captures patent records from 1996-2014, patents applied in latter years that are granted after 2014 are not included as granted. Thus, compared to USPTO national data, the aggregate of patents granted in our data set tapers after 2011 (Figure 3). We have estimated our results without observations in 2012-2014 and found no significant difference. We therefore retain all years of observations despite the discrepancy. 

\vspace{.3cm}
\begin{figure}[h]
\centering
\caption{}
\includegraphics[width=14cm]{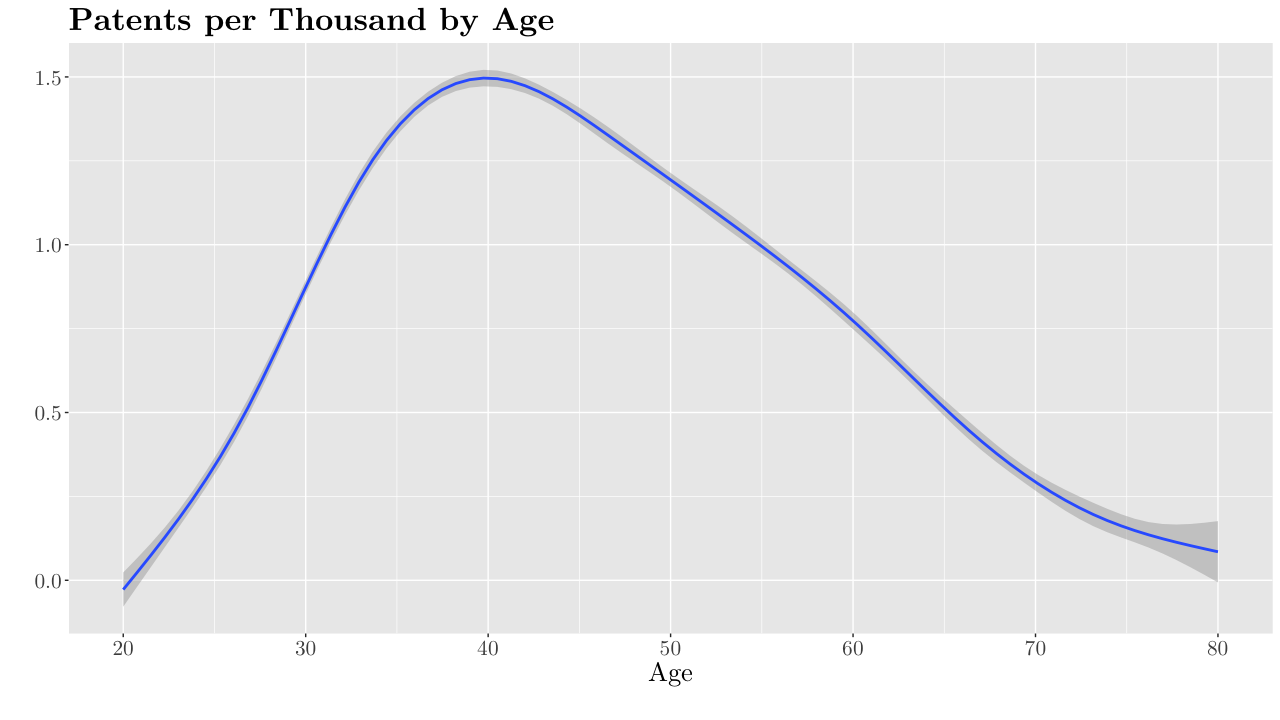}
\end{figure}

\vspace{.3cm}
\begin{figure}[h]
\centering
\caption{}
\includegraphics[width=15cm]{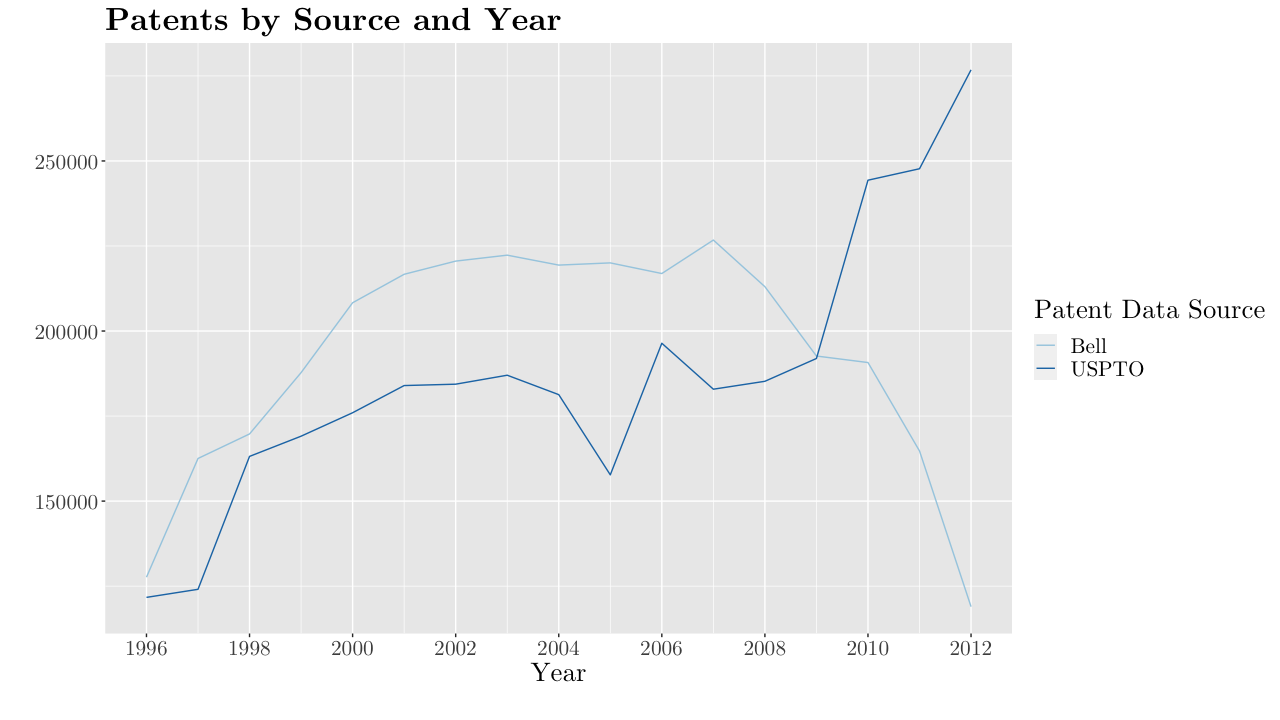}
\end{figure}

Table 1 reports summary statistics. Our outcome of interest is patents granted per thousand people by cohort, state, and year. The mean of the outcome is 0.887. We use patents adjusted by population as an outcome as a proxy for research productivity. For two groups with a given number of people, if group A is granted more patents than B, then group A has higher research productivity. Other variables of interest from the data set include each cohort’s estimated population in each year in a given state. If having a larger cohort population at birth causes the cohort to later create more patents, all else equal, then a larger population base would be causing higher research productivity. The mean count of residents by cohort, state is 60,075. \vspace{.3cm}

\begin{table}[!htbp] \centering 
  \caption{Descriptive Statistics by Cohort, State, and Year (unless otherwise specified)} 
  \label{Descriptive Statistics by Cohort, State, and Year (unless otherwise specified)} 
\begin{tabular}{@{\extracolsep{0pt}}lccccc} 
\\[-1.8ex]\hline 
\hline \\[-1.8ex] 
Statistic & \multicolumn{1}{c}{N} & \multicolumn{1}{c}{Mean} & \multicolumn{1}{c}{St. Dev.} & \multicolumn{1}{c}{Min} & \multicolumn{1}{c}{Max} \\ 
\hline \\[-1.8ex] 
Patents granted per 1,000 & 41,578 & 0.887 & 0.976 & 0.000 & 12.751 \\ 
Age & 41,578 & 47.390 & 15.395 & 20 & 80 \\ 
Population by Year & 41,578 & 60,075.030 & 71,862.380 & 1,710 & 504,530 \\ 
Births by Cohort, State & 41,578 & 64,779.690 & 70,608.410 & 1,223 & 998,198 \\ 
Per Capita Income by Cohort, State & 41,578 & 3,032.193 & 3,060.466 & 122 & 19,701 \\ 
\hline \\[-1.8ex] 
\end{tabular} 
\end{table}

We then combine patent outcomes with birth statistics tracked by the CDC’s Vital Statistics of the United States reports. Births are reported by state and by cohort. On average, 64,780 children are given birth to in each cohort by state. This provides us with an estimate of a cohort’s population at birth, which not only is essential to our estimation of population effects on patent production for each cohort, but also allows us to compare how much each cohort’s population has changed since birth. Because birth data is available starting in 1931, we limit our initial sample of cohorts accordingly. We also merge per capita personal income by state published by the Federal Reserve, so as to control for income’s impact on each state’s average reproductive decisions. 

\subsection{Model Specification}

We first estimate the following equation of patents per thousand people on the determinants of patent production:
\begin{align*}
    p_{i,s,t} = \text{Birth}_{i,s} \delta +  \text{Migrate}_{i,s,t}\beta + \text{Inc}_{i,s}\rho +  \text{Age}_{i,t}\alpha +  \text{Age}^2_{i,t}\lambda + \gamma_t + \epsilon_{i,s,t}
\end{align*}

Here, $p_{i,s,t}$ refers to patents granted per thousand people to an observation, which is a given cohort $i$ living in state $s$ in the year $t$. $\emph{Birth}_{i,s}$ refers to the number of births in a state for a given cohort year. $\emph{Migrate}_{i,s,t}$ refers to the ratio of current residents in an observation to $\emph{Birth}_{i,s}$. The way to interpret the ratio is that since the starting population for each cohort is fixed throughout its lifetime, differences between the current resident count and births is most likely attributable to migration, either among states or in and out of the US. This allows us to attempt to hold constant the effect of job-motivated relocations or high-skilled immigration on an observation’s patent production. 

$\emph{Inc}_{i,s}$ refers to a state’s per capita personal income recorded for the year during which a cohort is born. Controlling for an observation’s state per capita income at birth has two purposes: (i) it allows us to control for reproductive choice differences across states that are driven by the differences in the ability to afford childrearing and (ii) it can imprecisely control for differences in human capital investment driven by income that can lead to different patent production outcomes. Unfortunately, including more recent income would lead to over controlling, as a state’s per capita income in year $t$ is some function of labor and capital input and technology; controlling for per capita income in year $t$ would essentially be controlling for the outcome. The same applies to most time-variant economic variables, where $t > i$. 

Given that patents granted quadratically rise and fall with a cohort’s age, we include $\emph{Age}_{i,t}$ as a quadratic term to control for shortfalls in patents granted to earlier and later cohorts that are likely due to their being too old and young, respectively. Finally, $\gamma_{t}$ are year dummies. We do not add cohort or state fixed effects, as the variation in birth levels are \emph{across cohorts and states}. Controlling for cohorts, i.e. looking at variation within cohorts, would not be sensible. The same applies to state fixed effects.

\subsection{OLS Results}

We estimate the aforementioned model and its simpler variations using OLS (see Table 2). We estimate heteroskedasticity-robust standard errors. A cohort’s starting population is significantly and positively correlated with patents granted to that cohort. The significant and positive relationship of births is robust across various specifications. Cohorts with larger populations tend to be granted more patents per capita. Using the full model’s coefficient, increasing births by one standard deviation (70,608) predicts an increase of 0.15 in patents per thousand (or 15 percent of the outcome’s standard deviation). This is of course not causal: this relationship may be driven by a co-moving variable increasing both patent production and births, such as parents’ research productivity. Another explanation could be that younger cohorts simply produce less patents on one hand, while younger cohorts are also smaller in population due to exogenous declines in birth nationally on the other. This would make births and outcomes correlated by coincidence. To infer causality, we turn to exogenous birth shocks in the next section.

\vspace{.3cm}

\begin{table}[!htbp] \centering 
  \caption{Relationship Between Patents Granted per Capita and Birth} 
  \label{Relationship Between Patents Granted per Capita and Birth} 
\begin{tabular}{@{\extracolsep{0pt}}lccc} 
\\[-1.8ex]\hline 
\hline \\[-1.8ex] 
 & \multicolumn{3}{c}{\textit{Dependent variable:}} \\ 
\cline{2-4} 
\\[-1.8ex] & \multicolumn{3}{c}{Patents per 1,000} \\ 
 & \multicolumn{3}{c}{OLS}  
\\[-1.8ex] & (1) & (2) & (3)\\ 
\hline \\[-1.8ex] 
 Births & 0.000002$^{***}$ & 0.000002$^{***}$ & 0.000002$^{***}$ \\ 
  & (0.0000001) & (0.0000001) & (0.0000001) \\ 
  Migration &  & 0.10$^{***}$ & 0.10$^{***}$ \\ 
  &  & (0.005) & (0.005) \\ 
  State Per-Capita Income at Birth &  & $-$0.00003$^{***}$ & $-$0.00004$^{***}$ \\ 
  &  & (0.000002) & (0.000002) \\ 
 \hline \\[-1.8ex] 
Year Fixed Effects & No & No & Yes \\ 
Age Controls & Yes & Yes & Yes \\ 
Heteroskedasticity-robust Wald stat & 5288.7*** & 3270.6*** & 771.47*** \\ 
Observations & 41,578 & 41,578 & 41,578 \\ 
R$^{2}$ & 0.20 & 0.20 & 0.22 \\ 
Adjusted R$^{2}$ & 0.20 & 0.20 & 0.22 \\ 
Residual Std. Error & 0.88 (df = 41574) & 0.87 (df = 41572) & 0.86 (df = 41556) \\ 
\hline 
\hline \\[-1.8ex] 
\multicolumn{4}{r}{$^{*}$p$<$0.1; $^{**}$p$<$0.05; $^{***}$p$<$0.01} \\ 
\end{tabular} 
\end{table}

We should note a few more things from the basic estimation. As expected, the more a cohort’s population swells as a result of migration, the more likely it is that it receives more patent grants per person. This could either be due to the network effects of high population, or the states having high research productivity attracting more residents. That state’s per capita income at birth is significantly negative on patents granted is surprising, though the coefficient relatively lacks economic insignificance. One possible reason is that holding a cohort’s starting population and migration fixed, income of the state from the cohort year has little impact on that cohort’s present patent production. Another reason is likely data structure. On one hand, state per capita income is higher in the years of birth of younger cohorts. On the other, younger cohorts also have less lifetime patent production between 1996-2012 due to their being younger. \vspace{.5cm}

\section{Causal Inference with Abortion Law as a Fertility Shock}
\subsection{Institutional Setting of Abortion Law}

In January of 1973, the Supreme Court of the United States ruled in the case \emph{Roe v. Wade} that women have a right to seek an abortion without state interference in the first trimester of their pregnancy. Before this ruling, around 1970, approximately two-fifths of states had already undergone liberalization reforms to their abortion laws as part of a series of criminal law reforms recommended by the legal profession at the time. We take advantage of the ruling as a heterogeneous negative shock on births which impacted states which had yet to undergo reforms, more than states where abortion law was already reformed. We assume both the pre-\emph{Roe} reforms and the \emph{Roe} ruling itself were sufficiently exogenous. In this sub-section, we briefly describe the state of abortion law before the \emph{Roe} ruling to justify our assumption.

According to Boonstra et al. (2006) from the Guttmacher Institute, there were 17 states which had liberalized abortion laws in some form of another. Before these reforms, abortions were generally criminalized at all stages of pregnancy across states (Buell 1991, 1787). 13 of the 17 states expanded exceptions which made abortions permissible; under the reforms, pregnancies could generally be terminated if they threatened the mother’s health, could result in a child with severe disability, or were the consequence of rape or incest (Tyler 1983, 245). The remaining four states decriminalized abortions and made them available to women who requested them (subject to physician approval), regardless of justification (Buell 1991, 1798-9; Tyler 1983, 247). Table 3 shows the dates of reforms for which there are records.
\vspace{.3cm}


\begin{table}[!htbp] \centering 
  \caption{Timing and Type of Abortion Law Reforms by State}
  \label{Timing and Type of Abortion Law Reforms by State} 
\begin{tabular}{@{\extracolsep{0pt}} cccc} 
\\ \hline 
\hline  \\ 
 & State & Reform\_Year & Reform \\ 
\hline 
1 & Alaska & $1970$ & Decriminalization \\ 
2 & Arkansas & $1969$ & Expanded exceptions \\ 
3 & California & $1967$ & Expanded exceptions  \\ 
4 & Colorado & $1967$ & Expanded exceptions  \\ 
5 & Delaware & $ $ & Expanded exceptions  \\ 
6 & Florida & $ $ & Expanded exceptions  \\ 
7 & Georgia & $ $ & Expanded exceptions  \\ 
8 & Hawaii & $1970$ & Decriminalization \\ 
9 & Kansas & $ $ & Expanded exceptions \\ 
10 & Maryland & $ $ & Expanded exceptions  \\ 
11 & New Mexico & $1969$ & Expanded exceptions  \\ 
12 & New York & $1965$ & Expanded exceptions; decriminalization in 1970 \\ 
13 & North Carolina & $1969$ & Expanded exceptions  \\ 
14 & Oregon & $1970$ & Expanded exceptions  \\ 
15 & South Carolina & $$ & Expanded exceptions  \\ 
16 & Virgina & $$ & Expanded exceptions  \\ 
17 & Washington & $1970$ & Expanded exceptions; decriminalization in 1971 \\ 
\hline 
\end{tabular} 
\end{table} 

Notes: The table combines information on 
          abortion law before Roe from Boonstra et al. (2006), 
          Buell (1991), Gold (2003), Milman (1970), and Tyler (1983).

We determine that the reforms were exogenous because they were initiated by the American Law Institute, which was publishing a Model Penal Code, a \emph{series} of criminal law reforms, one of whose many subjects included abortion law (Buell 1991, 1796; Tyler 1983, 245; Boonstra et al. 2006, 12). The Model Penal Code was influential and generally adopted by states without partisan disagreement. To this end, we estimate a probability model of a state having bans on abortion (i.e. not having undergone reform) with logistic regression. The vote share data is obtained from UC Santa Barbara's American Presidency Project. The results are in Table 4. The status of having undergone reform is not predictable by voting behavior. Although state per capita income in 1971 as a predictor has moderate statistical significance, it is not economically significant. 

\vspace{.3cm}

\begin{table}[!htbp] \centering 
  \caption{Determinants of Abortion Bans} 
  \label{Determinants of Abortion Bans} 
\begin{tabular}{@{\extracolsep{0pt}}lc} 
\\[-1.8ex]\hline 
\hline \\[-1.8ex] 
 & \multicolumn{1}{c}{\textit{Dependent variable:}} \\ 
\cline{2-2} 
\\[-1.8ex] & Probability of State Keeping Abortion Banned \\ 
 & Logit \\ 
\hline \\[-1.8ex] 
 Vote Share for Rep., 1968 Presidential & $-$6.467 \\ 
  & (7.927) \\ 
  Vote Share for 3rd Parties, 1968 Presidential & $-$8.993 \\ 
  & (7.290) \\ 
  Vote Share for Rep., 1972 Presidential & $-$0.571 \\ 
  & (8.902) \\ 
  State Per-Capita Income in 1970 & $-$0.002$^{**}$ \\ 
  & (0.001) \\ 
  Constant & 12.396$^{**}$ \\ 
  & (5.957) \\ 
 \hline \\[-1.8ex] 
Observations & 50 \\ 
Log Likelihood & $-$27.724 \\ 
Akaike Inf. Crit. & 65.449 \\ 
\hline 
\hline \\[-1.8ex] 
\multicolumn{2}{r}{$^{*}$p$<$0.1; $^{**}$p$<$0.05; $^{***}$p$<$0.01} \\ 
\end{tabular} 
\end{table} 

Separately, the Supreme Court’s ruling expanded abortion access far beyond what was legislatively achieved, as it declared that the right to privacy included the right to an abortion and left the judgement of abortion’s suitability during the first trimester solely to the physician. A latter ruling found that features shared by most reformed laws pre-\emph{Roe} violated this right, e.g. the requirement of two physicians jointly approving the abortion (Tyler 1983, 248; 249). Furthermore, an overwhelming majority of public opinion as of 1969 was opposed to permitting abortions for households which would have financial difficulties raising children or for those which simply did not prefer additional children (Blake 1971, 541). For this reason, the liberalization of abortion by \emph{Roe} appeared to be the result of jurisprudence that was uncorrelated with the pre-\emph{Roe} appetite for legislative reforms or public opinion at the time. 

Levine et al. (1999) estimate that before the decision in 1973, states which allowed abortions saw 4 percent declines in fertility after abortion was allowed relative to states where it remained illegal; prior to \emph{Roe v. Wade}, Levine et al. find evidence that women travelled out of state to obtain abortions (quoted in Lopoo and Raissian 2012, 928-930). We should note that out-of-state abortions are not a problem for our data set of births, as we only track births in each state. A case where a woman–who would have sought abortion out-of-state pre-\emph{Roe}–requesting an abortion in her home state after \emph{Roe} would make no difference to the record of births in her home state or the destination state where the abortion would have been sought (as the birth would not have occurred and therefore not changed the birth record in either case). Our birth data set should therefore accurately track change in births pre- and post-\emph{Roe}.

Contextually, the period surrounding the \emph{Roe} as a negative exogenous shock occurs during a time where numerous pro and anti-birth policies were enacted. In Figure 4, we plot the trend of US birth per thousands and year-on-year percentage change in population, in addition to highlighting when key birth policies occurred (pro-birth dotted in blue, anti-birth in red, ambiguous in grey).  We see from Figure 4 that US birth rates have broadly been trending down since the introduction of the birth control pill in 1960, with Bailey (2010) estimating that 40 percent of the decline in fertility from 1955 to 1965 was attributable to the pill (quoted in Lopoo and Raissian 2012, 931). This trend was heightened by the introduction of the national Publicly Funded Family Planning (PFFP) Act in 1964, which provided “access to contraception, counselling services, preventive care, and health screenings throughout the country, with a particular emphasis on the low-income population” (Lopoo and Raissian 2012, 932-933). Gold (2001) estimates that some 20 million pregnancies were averted between 1980 to 1999 as a result (quoted in Lopoo and Raissian 2012, 933).

\vspace{.3cm}
\begin{figure}[h]
\centering
\caption{}
\vspace{.5cm}
\includegraphics[width=17.5cm]{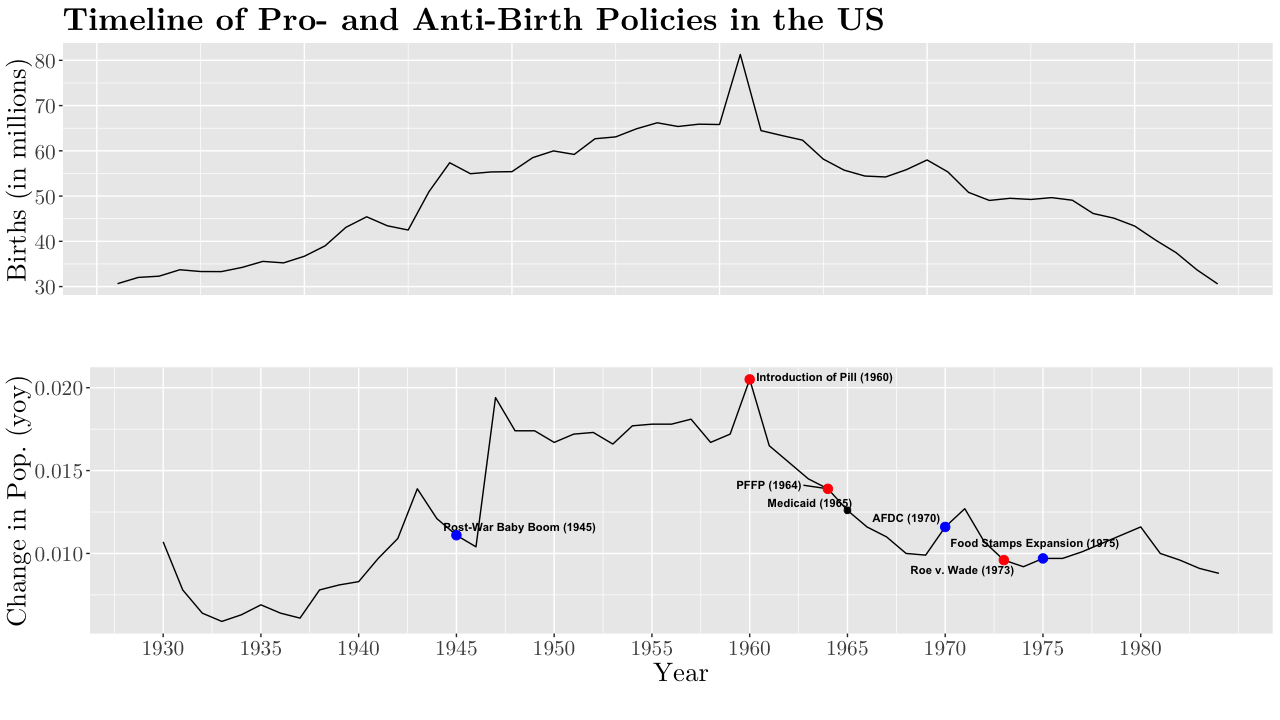}
\end{figure}

There were pro-birth policies too: (1) the Aid to Families Dependent Children (AFDC) Act which was introduced in 1935 but expanded in 1970; (2) Medicaid was introduced in 1965; and (3) the Food Stamps / Supplemental Nutrition Assistance Program (FSP / SNAP) were piloted in 1961 and expanded in 1975. These programs all in some form provided a form of cash or indirect subsidy for families in need (Lopoo and Raissian 2012, 922-933). These non-trivial benefits of either cash or goods were disproportionally beneficial to low-income groups and likely increased probabilities of marriage and birth (Lopoo and Raissian 2012, 906; 911). Medicaid, for example, provided free medical insurance to low-income individuals and increased fertility rates by 29 percent among recipients (Lopoo and Raissian 2012, 921). Given that these policies occurred at the national level, we believe the subsequent trends can generally be controlled for with a comparable control group of states, hence our use of DID.

\subsection{Specification: DID}

We estimate the DID of patents per thousand with the following model:
\begin{align*}
    p_{i,s,t} = (\text{Ban}_{s} \times \text{Roe}_{i}) \theta_{1} + \text{Ban}_{s}\theta_2 + \text{Roe}_{i}\theta_3 +  \text{Migrate}_{i,s,t}\beta + \text{Inc}_{i,s}\rho +  \text{Age}_{i,t}\alpha +  \text{Age}^2_{i,t}\lambda + \gamma_t + \epsilon_{i,s,t}
\end{align*}

The model is the same as our original model, except that $\emph{Birth}s$ is replaced by an interaction term of two dummies, where $\emph{Ban}_{s} = 1$ for states $s$ which banned abortions up until Roe, and $\emph{Roe}_i \space \forall \space i \geq 1973$. 

Separately we also use the interaction term as an instrumental variable, with which we estimate the following first-stage equation of a two-stage least squares (2SLS) estimation:
\begin{align*}
    \text{Births}_{i,s} = (\text{Ban}_{s} \times \text{Roe}_{i}) \ell_{1} + \text{Ban}_{s}\ell_2 + \text{Roe}_{i}\ell_3 + \eta_{i,s}
\end{align*}
The resulting $\widehat{Births}_{i,s}$ is used to to estimate the original model. As it will become clear, we felt it was necessary to isolate the impact of \emph{Roe} on cohort starting sizes specifically. This is the case because as we will see, even if the two groups of states followed the parallel trends before \emph{Roe}, we think it is possible that they might have diverged after \emph{Roe}. This likely would have been the case due to the cumulative effect of policy changes that occurred at the state, and not to mention county and municipal level, after \emph{Roe}.  \vspace{.5cm}

\section{Results}

\subsection{DID}
Column 1 of Table 5 displays the results from the vanilla difference-in-difference model. Contrary to our expectation, a cohort born in a state with pre-\emph{Roe} abortion bans (which should have experienced a negative shock in fertility) were not granted less patents per thousand people later in their lifetime. In fact, the estimated coefficient is weakly positive, which suggests that non-reform states on average, compared to reform states, saw slightly greater positive change in patents per capita after \emph{Roe}. 
\vspace{.3cm}

\begin{table}[!htbp] \centering 
  \caption{Estimating Patents Granted per Capita on Births using Roe v. Wade and Pre-Roe Bans} 
  \label{Estimating Patents Granted per Capita on Births using Roe v. Wade and Pre-Roe Bans} 
\begin{tabular}{@{\extracolsep{0pt}}lccc} 
\\[-1.8ex]\hline 
\hline \\[-1.8ex] 
 & \multicolumn{3}{c}{\textit{Dependent variable:}} \\ 
\cline{2-4} 
\\[-1.8ex] & Patents per 1,000 & Births & Patents per 1,000 \\ 
 & DID & 2SLS First-stage & 2SLS Second-stage \\ 
\\[-1.8ex] & (1) & (2) & (3)\\ 
\hline \\[-1.8ex] 
 Post-Roe × Pre-Roe Ban & 0.12$^{**}$ & $-$2,675.00 &  \\ 
  & (0.06) & (4,346.54) &  \\ 
  Births &  &  & 0.000003$^{***}$ \\ 
  &  &  & (0.000001) \\ 
  Migration & 0.19$^{***}$ &  & 0.23$^{***}$ \\ 
  & (0.03) &  & (0.03) \\ 
  State Per-Capita Income at Birth & 0.00005$^{***}$ &  & 0.00002$^{***}$ \\ 
  & (0.000005) &  & (0.00001) \\ 
 \hline \\[-1.8ex] 
Year Fixed Effects & Yes & No & Yes \\ 
Age Controls & Yes & No & Yes \\ 
Heteroskedasticity-robust Wald stat &  & 51.971*** (df = 3; 9692) &  \\ 
Observations & 9,696 & 9,696 & 9,696 \\ 
R$^{2}$ & 0.29 & 0.02 & 0.29 \\ 
Adjusted R$^{2}$ & 0.29 & 0.02 & 0.29 \\ 
Residual Std. Error & 0.69 (df = 9672) & 69,035.17 (df = 9692) & 0.69 (df = 9674) \\ 
\hline 
\hline \\[-1.8ex] 
\multicolumn{4}{r}{$^{*}$p$<$0.1; $^{**}$p$<$0.05; $^{***}$p$<$0.01} \\ 
\end{tabular} 
\end{table}

To understand these results better, Figure 5 plots mean of patents per capita over states and years in each group. Mean patents per capita over years here is the prediction–from estimating said outcome as a quadratic function of cohort’s mean age–where every cohort is assumed to be age 40, plus the residual from said estimation. The outcome is then averaged across the control and treatment groups, respectively. 

\vspace{.3cm}
\begin{figure}[h]
\centering
\caption{}
\includegraphics[width=15cm]{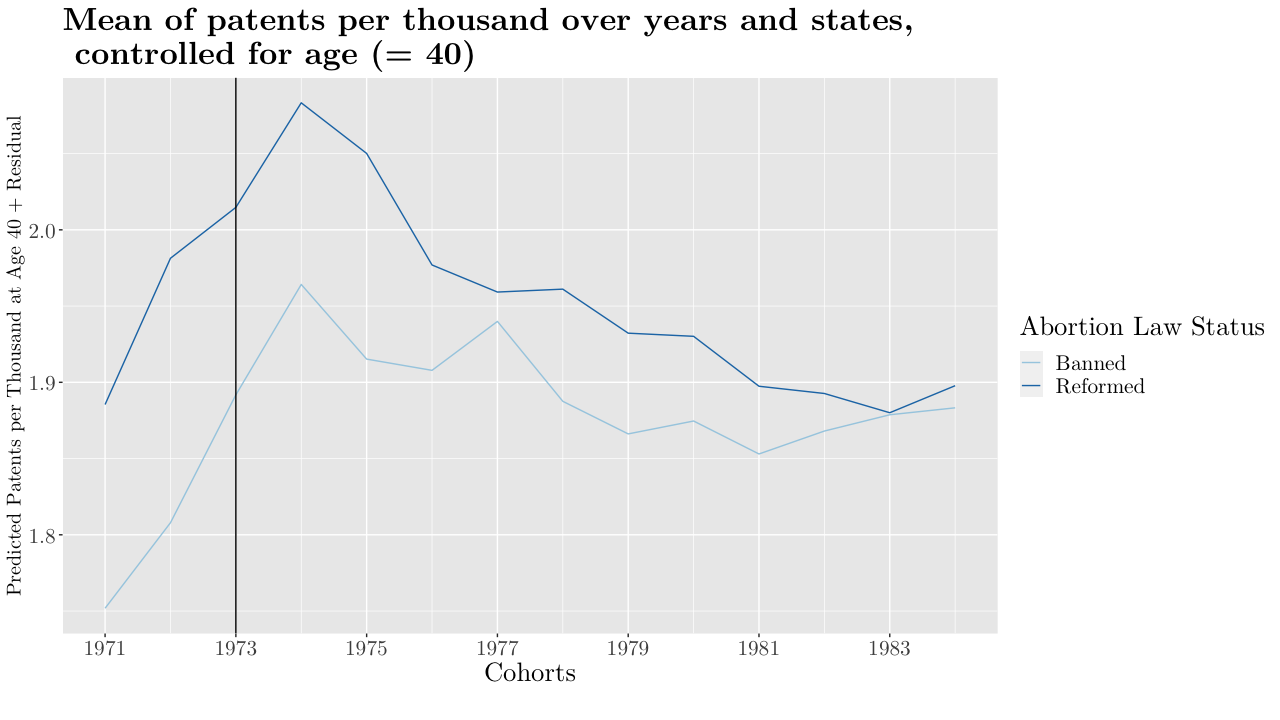}
\end{figure}

We observe two things: first, parallel trends held in the first two years. The optimistic interpretation is that the effect of pre-\emph{Roe} reforms (mostly from before 1970) had already realized their effects on pre-1971 cohorts, and therefore we see cohorts from reformed states following the same national trends as cohorts from non-reformed states. The pessimistic interpretation is that those reforms had yet to realize their impact and did not until \emph{after Roe}, which would violate parallel trends. Assuming parallel trends of births hold for now, the second observation is that not only did non-reform states (i.e. the states “treated” by the \emph{Roe} shock) not see a drop in patent outcomes, they actually saw a rise in patent per capita outcomes relative to reform states. This would explain the positive interaction coefficient. This could be explained by the violation of parallel trends more broadly, and more specifically by non-birth factors that occurred predominantly to one of two groups after \emph{Roe}.

Non-birth dynamics after \emph{Roe} violating parallel trends may explain why \emph{Roe} did not have the intended effect on patents per capita. Our dataset spans multiple decades, and this duration inherently increases the likelihood of being unable to satisfy the parallel trend assumption critical to an unbiased DID. The period before and after \emph{Roe} saw the patent industry transform with the economy during the 1990s and 2000s due to to the internet, which expanded the number of invention types which were deemed patentable (OECD 2004, 5-7); the patent industry predictably grew exponentially. Software patents, for example, grew fourfold from 1990 to 2000 and comprised 15 percent of all patents granted in the US in 2000, outpacing growth for the rest of the patent industry which was short of doubling over the same period (OECD 2004, 7; 24). The patent industry’s pre-\emph{Roe} trajectory may have been disrupted entirely and with it the parallel trends between the two groups of states. While one could argue that the internet era generated equally positive impacts on patent production across all states, which is what our model implicitly assumed. But it is very likely that such a transformation was not equal for all states. As we elaborate further below, other factors relating fundamentally to how patents are issued and who applies for patents, along with other potential non-parallel trends, could have compounded the effects of an unequal distribution of growth in the patent industry and skewed results.

The role of companies and tax jurisdiction may also have played sizeable roles in disrupting state dynamics in terms of where patents are allocated. Research has shown that business-financed R\&D in OECD countries (including the US) rose 51 percent in real terms between 1990 and 2001 (OECD 2004, 15), hinting that the exponential growth of patent issuance during this period was largely driven by corporates.\footnote{Note that pre-invention assignments (agreements between employers and employees to assign invention rights to employers) meant patents can and were frequently allocated to employers rather than individuals/employees.} This is one of the areas in which the patent industry has evolved from pre- to post-\emph{Roe}, which could have caused our parallel trends assumption to fail. For example, tax jurisdiction plays a major role in where companies incorporate and operate, and companies' flexibility for relocating their headquarters for tax purposes may have unobservable effects on which states patents end up being assigned to. There are two ways of this happening that can violate parallel trends and explain the relative rise in patents per capita in non-reform states even if population has a positive impact on patent production. First, non-reform states might have attracted companies (consisting of employees from younger cohorts) with lower corporate tax rates and therefore the origin of their patents from reform states to their non-reform counterparts. Second, as Saez and Zucman (2019) document, profit-shifting through the sale of patents to foreign subsidiaries has been rampant, especially among tech companies (which hire younger cohorts) (Saez and Zucman 2019, 71-81). If the same motivations also disproportionately shift the registration of patents from reform states to abroad, this would violate parallel trends and could explain the positive DID coefficient. Overall, the changing landscape of the patent industry is difficult to fully control for, which makes our assumption of stable trends across the two groups of states difficult to hold.

We now examine some of the core demographic covariates raised in Bell et al. and other research, which also may have played a role in violating parallel trends. Recall that Bell et al.’s paper found factors such as gender, race and ethnicity to be significant for explaining innovation. We have simply assumed that they satisfy parallel trends. Similarly, spatial dimension factors such as proximity to economic centers were also assumed parallel. Again, given the duration of our model, assumptions on the stability of financial conditions, commuting conditions and levels of exposure to innovation may have been too optimistic. Our model could not control for these factors due to lack of data on commuting and gender by cohort and state. Future research could unpack how these dynamics have evolved differently across states over time, and whether they have significant impacts on patent production across states. 

\subsection{IV-DID}

The previous section showed that parallel trends at-large likely did not hold in the post-\emph{Roe} period. For this reason, we need an approach that isolates the effect of the \emph{Roe} ruling on births exclusively. We therefore turn to the results from using difference-in-difference after \emph{Roe} as an instrumental variable (IV-DID) for estimating births as the treatment variable in a 2SLS setup. This would hopefully allow us to limit the scope of the parallel trends assumption to births only. Assuming the assumption holds, we can obtain a causal estimate of population effects on innovation from this setup.

The results are also shown in Table 5. Column 2 shows that the DID interaction term has a negative sign as we expect, but lacks statistical significance. The heteroskedasticity-robust Wald test statistic of the first-stage regression suggests that our IVs are not weak. However, most of the statistical significance of the predictors can be attributed to the dummy for classifying reform and non-reform states, rather than the interaction term. To understand the strength of the \emph{Roe} IVs better, we plot births by cohorts and abortion law status in Figure 6. 

\vspace{.3cm}
\begin{figure}[h]
\centering
\caption{}
\vspace{.5cm}
\includegraphics[width=15cm]{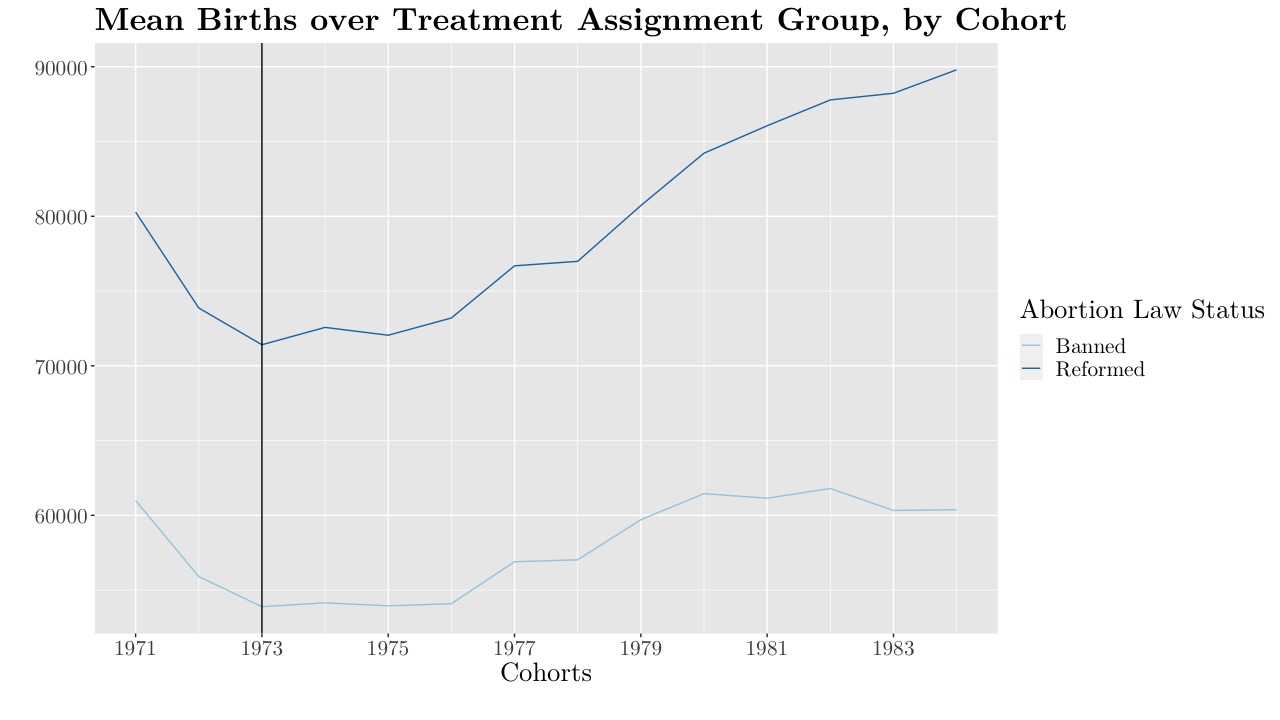}
\end{figure}

First, we can immediately see that reform states began with a higher level of births, which explains the strength of the \emph{Roe} IVs as a whole at predicting births. Second, we also see that non-reform states did experience less births after \emph{Roe}, as we hypothesized. Third, birth trends in non-reform and reform states for cohorts after \emph{Roe} did not diverge until after 1978.  This might be attributable to reforms happening too close to the \emph{Roe} ruling and their delayed impact, as we discussed earlier in our diagnosis of the DID results. Even if the pre-\emph{Roe} abortion law reforms were exogenous, if their negative effects on birth did not fully realize until after \emph{Roe}, then \emph{Roe}’s effect on the non-reform states woukd be underestimated (as the delayed effect of the pre-\emph{Roe} reforms in the reform states would have driven up the average number of births pre-\emph{Roe}, leaving the change in births in reform states greater, and therefore the change in births in non-reform states relatively smaller). We experiment with shifting the post-\emph{Roe} cutoff to 1978, and the interaction term’s heteroskedasticity-robust $t$-value in the first-stage regression does indeed become stronger, going from $-0.62$ to $-1.56$.

Our discussion from the previous section suggests that our IV is likely to underestimate the negative effect of births on non-reform states. Nonetheless, the second-stage regression estimates larger cohorts to have a significant and positive effect on patents outcome. One standard deviation increase in births is predicted to increase patents per capita by 0.23 (or 24 percent of one standard deviation of the outcome). Another indicator suggesting a causal estimate is that the coefficient for state per capita income at a cohort’s birth is positive, which aligns with our expectation (i.e. richer cohorts produce more patents due to exposure to parents with high research productivity or higher levels of human capital investment) and suggests that if the effect of income were captured by births in the basic estimation (which would bias our estimate of population's effect), this is no longer the case.

\subsection{Robustness}

Table 6 considers an alternative specification of the DID term. Specifically, there are two interaction terms:
\begin{align*}
    \text{Births}_{i,s} = (\text{Ban}_{s} \times \text{Roe}_{i}) \kappa_1 + (\text{Restricted}_{s} \times \text{Roe}_{i})\kappa_2 + \text{Ban}_{s}\kappa_3 + \text{Restricted}_{s}\kappa_4 + \text{Roe}_{i}\kappa_5 + \upsilon_{i,s}
\end{align*}

\begin{table}[!htbp] \centering 
  \caption{Estimating Patents Granted per Capita on Births using Roe v. Wade and Pre-Roe Bans and Restrictions} 
  \label{Estimating Patents Granted per Capita on Births using Roe v. Wade and Pre-Roe Bans and Restrictions} 
\begin{tabular}{@{\extracolsep{0pt}}lccc} 
\\[-1.8ex]\hline 
\hline \\[-1.8ex] 
 & \multicolumn{3}{c}{\textit{Dependent variable:}} \\ 
\cline{2-4} 
\\[-1.8ex] & Patents per 1,000 & Births & Patents per 1,000 \\ 
 & DID & 2SLS First-stage & 2SLS Second-stage \\ 
\\[-1.8ex] & (1) & (2) & (3)\\ 
\hline \\[-1.8ex] 
 Post-Roe × Pre-Roe Ban & 0.11$^{**}$ & $-$4,627.68 &  \\ 
  & (0.05) & (4,511.29) &  \\ 
  Post-Roe × Pre-Roe Restriction & 0.03 & 8,298.90 &  \\ 
  & (0.16) & (10,757.15) &  \\ 
  Births &  &  & 0.000002$^{***}$ \\ 
  &  &  & (0.000001) \\ 
  Migration & 0.20$^{***}$ &  & 0.22$^{***}$ \\ 
  & (0.03) &  & (0.03) \\ 
  State Per-Capita Income at Birth & 0.00004$^{***}$ &  & 0.00003$^{***}$ \\ 
  & (0.00001) &  & (0.00001) \\ 
 \hline \\[-1.8ex] 
Year Fixed Effects & Yes & No & Yes \\ 
Age Controls & Yes & No & Yes \\ 
Heteroskedasticity-robust Wald stat &  & 31.226*** (df = 5; 9690) &  \\ 
Observations & 9,696 & 9,696 & 9,696 \\ 
R$^{2}$ & 0.29 & 0.02 & 0.31 \\ 
Adjusted R$^{2}$ & 0.29 & 0.02 & 0.31 \\ 
Residual Std. Error & 0.69 (df = 9670) & 69,026.57 (df = 9690) & 0.68 (df = 9674) \\ 
\hline 
\hline \\[-1.8ex] 
\multicolumn{4}{r}{$^{*}$p$<$0.1; $^{**}$p$<$0.05; $^{***}$p$<$0.01} \\ 
\end{tabular} 
\end{table} 

Here, $\emph{Restricted}_{s} = 1$ for both non-reform states and 13 of the 17 states which have undergone reforms but not decriminalized abortion completely. The idea is that if each state’s exposure to the Roe shock is a function of abortion law stringency, then both parameters should be negative. Table 6 shows the results. The 2SLS second-stage regression results (Column 3) are consistent with the previous estimation; births have a significantly positive effect on patents per thousand. Meanwhile, the DID interaction terms are still positive, which is consistent with the previous estimates and similarly contradict our hypothesis.

Column 2 of Table 6 shows how well the two dummies predicts births in the first-stage regression. Though neither terms are significant, contrary to our expectation, the interaction term for restrictions is positive, meaning that the group of states had on average experienced a greater, positive change in births after \emph{Roe}, than the four states (Alaska, Hawaii, New York, and Washington) that decriminalized abortion. Compared to reform states which kept restrictions, non-reform states nonetheless on average had less births. 

This suggests to us that there is no relevant distinction between reform states which expanded the justifications with which women can permissibly seek abortion versus those which decriminalized abortion completely. If this is true, this can violate the exogeneity of \emph{Roe} as an IV, because this suggests that reform states are united in their similarity by some underlying variable despite different levels of abortion law stringency. It could be this same underlying variable that are driving the slight difference we see in births between reform and non-reform states, rather than the status of pre-\emph{Roe} abortion laws.
\vspace{.5cm}

\section{Conclusion}
This paper has presented evidence on how a larger population base affects technological growth by using the US Supreme Court's decision to legalize abortion for the first trimester nationwide in 1973 as a negative shock on births which disproportionately affected states which had not relaxed abortion laws prior to the ruling. To justify both the ruling and the status of pre-\emph{Roe} reforms as exogenous, we showed that states adopted reforms to their abortion law as one part in a series of criminal law reforms suggested by legal professionals in the 1960s, and that the Supreme Court's ruling was more far-reaching than most of these reforms or public opinion at the time. Using DID of the ruling as an IV for births, we have found that increasing births by one standard deviation increases patents per capita by 0.24 standard deviation. The results are consistent with Michael Kremer's theory that a larger population base causes higher levels of technological growth. We have analyzed why our instrumental variable lacks statistical significance and attribute this to the effect of pre-\emph{Roe} reforms being realized in a delayed manner in the period which coincided with the \emph{Roe} ruling's shock on births, leading to an underestimation of \emph{Roe's} relative impact on births in non-reform states. We have also analyzed why our simple DID estimates have returned estimates with signs contrary to our hypothesis. The two possible explanations we have identified are (i) shifts in intellectual property investment and corporate taxation and (ii) demographic shifts–for neither of which we have controlled.

\pagebreak


\begin{thebibliography}{}


\bibitem{}
Philippe Aghion and Peter Howitt. “A Model of Growth Through Creative Destruction." \emph{Econometrica} 60.2 (1992): pp. 323-351. 
\bibitem{}

Robert J. Barro and Gary S. Becker. “Fertility Choice in a Model of Economic Growth." \emph{Econometrica} 57.2 (1998): pp: 481-501

\bibitem{}
Alex Bell, Raj Chetty, Xavier Jaravel, Neviana Petkova, and John Van Reenen. “Who Becomes an Inventor in America? The Importance of Exposure to Innovation." \emph{Quarterly Journal of Economics} 134.2 (2019): pp. 647-713.

\bibitem{}
Judith Blake. “Abortion and the Public Opinion: The 1960-1970 Decade." \emph{American Association for the Advancement of Science} 171.3971 (1971): pp. 540-549.


\bibitem{}
Heather D. Boonstra, et al., \emph{Abortion in Women's Lives,} New York: The Guttmacher Institute, 2006. 

\bibitem{}
Samuel W. Buell. “Criminal Abortion Revisited." \emph{New York University Law Review} 66.1774 (1991): pp. 1774-1831. 

\bibitem{}
Center for Disease and Control (CDC). \emph{National Vital Statistics Report - Births: Final Data, 1931-1982}. Hyattsville, MD. Distributed by National Center for Health Statistics.  https://www.cdc.gov/nchs/nvss/births.htm

\bibitem{}
Oded Galor and David N. Weil. "From Malthusian Stagnation to Modern Growth." \emph{The American Economic Review} 89.2 (1999): pp. 150-154.

\bibitem{}
Rachel Benson Gold. “Lessons from Before \emph{Roe}: Will Past be Prologue?" \emph{The Guttmacher Report on Public Policy} (2003): 8-11.

\bibitem{}
Michael Kremer. “Population Growth and Technological Change: One Million B.C to 1990." \emph{The Quarterly Journal of Economics} 108.3 (1993): pp. 681-716. 

\bibitem{}
Simon Kuznets. “Population Change and Aggregate Output." In \emph{Demographic and Economic Change in Developed Countries}, edited by Universities-National Bureau, pp. 324-351. New York: Columbia University Press, 1960. 

\bibitem{}
Leonard M. Lopoo and Kerri M. Raissian. “Natalist Policies in the United States." \emph{Journal of Policy Analysis and Management} 31.4 (2012): pp. 905-946. 


\bibitem{}
Kathryn G. Milman. “Abortion Reform: History, Status, and Prognosis." \emph{Case Western Law Review} 21.3 (1979): 521-548.

\bibitem{}
 Organisation for Economic Co-operation and Development (OECD). \emph{Patents and Innovations: Trends and Policy Challenges.} Paris: Organisation for Economic Co-operation and Development, 2004.

\bibitem{}
Debraj Ray. \emph{Developement Economics.} New Jersey: Princeton University Press, 1998.

\bibitem{} 
Emmanual Saez and Gabriel Zucman. \emph{The Triumph of Injustice: How the Rich Dodge Taxes and How to Make Them Pay}. New York: W. W. Norton \& Company, 2019.


\bibitem{}
Carl W. Tyler Jr. “The Public Health Implications of Abortion." \emph{Annual Reviews} 4 (1983): pp.223-58.

\bibitem{}
U.S. Bureau of Economic Analysis and Federal Reserve Bank of St. Louis (FRED). \emph{State-Level Per Capita Personal Income, 1931-1982}.  St. Louis, MO, 13 Mar 2022. Distributed by FRED, Federal Reserve Bank of St. Louis. 

\bibitem{}
David N. Weil. “Why has Fertility Fallen Below Replacement in Industrial Nations, and Will it Last?" \emph{American Economic Association}  89.2 (1999): pp. 251-255. 

\bibitem{}
Stephen D Williamson. \emph{Macroeconomics, Fifth Edition.} New Jersey: Pearson, 2014. 

\bibitem{}
John Woolley and Gerherd Peters.  \emph{Statistics: ELections}. Produced and Distributed by The University of California Santa Barbara (UCSB) and The American Presidency Project. https://www.presidency.ucsb.edu/statistics/election

\bibitem{}
Hongyu Xiao, Andy Wu, and Jaeho Kim. “Commuting and innovation: Are closer inventors more productive?" \emph{Journal of Urban Economics} 121 (2021). 

\bibitem{}
Alwyn Young. “Invention and Bounded Learning by Doing." \emph{Journal of Political Economy} 101.3 (1993): pp. 443-472.



\end{thebibliography}
\end{document}